\documentclass[prb, preprint]{revtex4}
\usepackage{amssymb}
\usepackage{makeidx}
\usepackage{amsmath}
\usepackage{graphicx}
\usepackage{dcolumn}
\usepackage{bm}
\usepackage{epstopdf}
\usepackage{color}
\usepackage{txfonts}
\usepackage{microtype}

\begin{document}

\title{On the extreme nonlinear optics of graphene nanoribbons in the strong
coherent radiation fields}
\author{H. K. Avetissian}
\author{B. R. Avchyan}
\author{G. F. Mkrtchian}
\author{K. A. Sargsyan}

\affiliation{Centre of Strong Fields Physics, Yerevan State University, 0025, Yerevan,
Armenia}

\begin{abstract}
The generation of high-order harmonics in quasi-one-dimensional graphene
nanoribbons (GNRs) initiated by intense coherent radiation is investigated.
A microscopic theory describing the extreme nonlinear optical response of
GNRs is developed. The closed set of differential equations for the
single-particle density matrix at the GNR-strong laser field multiphoton
interaction is solved numerically. The obtained solutions indicate the
significance of the band gap width and Fermi energy level on the high-order
harmonic generation process in GNRs.
\end{abstract}

\date{\today }
\maketitle

\section{ Introduction}

Graphene and its analogs have attracted enormous interest in the last decade
due to the unique electronic and optical properties of such 2D quantum
systems \cite{Castro}. The significance of graphene as an effective
nonlinear optical material has triggered many theoretical \cite%
{H1,H2,H3,H4,H5,H6,H7,H8,H9,H10} and experimental \cite{HH2-exp,HH3-exp}
investigations devoted to diverse extreme nonlinear optical effects,
specifically, high-harmonic generation (HHG) taking place in the strong
coherent radiation fields -at the multiphoton excitation of such
nanostructures \cite{Mpc1,Mpc2}. On the other hand, apart from the
invaluable physical properties, two dimensional graphene can be patterned
into narrow ribbon that causes the carriers to be confined in
quasi-one-dimensional graphene nanoribbons (GNRs) (with the diverse
topologies depending on the ribbon form) \cite{Ribon}. Although the band
structure of a GNR differs for patterns with different boundaries, a common
feature of the GNRs is a width-dependent sizable band gap \cite{Brey}
suitable and significant for nano-opto-electronics. Such nanostructures
exhibit optical properties fundamentally different from those of graphene 
\cite{opt1,opt2,opt3}. At the same time, carriers in GNRs have the same
outstanding transport properties as in graphene \cite{Castro}.

The nonlinear optical response of graphene can be further enhanced via
plasmonic excitations supported by the graphene layer. Plasmons in\ graphene
can be manipulated by variation of the Fermi energy \cite{plasmon1}. At
that, graphene plasmons exhibit extreme subwavelength confinement \cite%
{plasmon2}. So that the strong near electric fields generated by plasmons in
graphene nanostructures can be exploited to enhance nonlinear optical
processes \cite{PLH1,PLH2,PLH3}. For extended graphene layer one can not
excite plasmons by a single wave field because of the energy-momentum
conservation law \cite{Constant}. Meantime, for patterned graphene
nanostructures this condition is vanished. Besides, plasmon frequencies can
be varied through the entire terahertz range \cite{Popov}. Hence, due to the
near field enhancement of the pump wave intensity one can realize the
extreme nonlinear regime of HHG when up to 100 harmonic orders can be
generated.

Another important advantage of GNRs over extended graphene monolayer is the
confinement of quasiparticles in GNRs in the one additional dimension. The
latter is crucial for HHG efficiency since confinement hinders the spread of
the electronic wave packet deposited to the continuum and, consequently,
enhances the HHG yield \cite{Corkum}. Hence, it is of interest to clear up
the influence of carrier confinement on the extreme nonlinear optical
response of GNRs, which is the subject of the current investigation.

In the present work, we develop a nonlinear microscopic theory of an
armchair GNR interaction with strong coherent electromagnetic (EM)
radiation. The theory of the interaction of confined carriers with a strong
driving wave-field is developed in the domain of the Dirac cone and
independent quasiparticles' approximations. The equation of motion for the
single-particle density matrix is solved numerically. Then we study the HHG
process in strong pump-waves and investigate HHG yield depending on the GNR
width size (dimers' number) and quasiparticles' Fermi energy level. Thus, we
predict high harmonics up to 80 orders in moderately strong pump
wave-fields/lasers.

The paper is organized as follows. In Sec. II the set of equations for the
single-particle density matrix is formulated. In Sec. III, we consider
multiphoton excitation of the Fermi-Dirac sea, and generation of harmonics
in GNR. Finally, conclusions are given in Sec. IV.

\section{Evolutionary equation for the single-particle density matrix}

Let an armchair GNR interacts with a plane quasimonochromatic\textrm{\ }EM
wave. We will consider an armchair GNR placed in the $XY$ plane bounded
along the $X$-axis and indefinite along the $Y$-axis. We assume that the
wave propagates in the perpendicular direction to the GNR plane. Thus, this
travelling wave for GNR electrons becomes a homogeneous quasiperiodic
electric field (of carrier frequency $\omega $ and slowly varying envelope $%
E_{0}\left( t\right) $). The polarization of the EM wave is assumed to be
parallel to the $Y$-axis: $\mathbf{E}\left( t\right) =\widehat{\mathbf{y}}%
E\left( t\right) $, where 
\begin{equation}
E\left( t\right) =E_{0}f\left( t\right) \cos \omega t.  \label{E_field}
\end{equation}%
The wave amplitude is described by the sine-squared envelope function $%
f\left( t\right) $:%
\begin{equation}
f\left( t\right) =\left\{ 
\begin{array}{cc}
\sin ^{2}\left( \pi t/\mathcal{T}_{p}\right) , & 0\leq t\leq \mathcal{T}_{p},
\\ 
0, & t<0,t>\mathcal{T}_{p},%
\end{array}%
\right.  \label{env}
\end{equation}%
where $\mathcal{T}_{p}$ characterizes the pulse duration.

Low-energy excitations which are much smaller than the nearest neighbor
hopping energy can be described by an effective Hamiltonian 
\begin{equation}
H_{0}=\hbar \mathrm{v}_{F}\,\left( 
\begin{array}{cccc}
0 & \widehat{k}_{x}-i\widehat{k}_{y} & 0 & 0 \\ 
\widehat{k}_{x}+i\widehat{k}_{y} & 0 & 0 & 0 \\ 
0 & 0 & 0 & -\widehat{k}_{x}-i\widehat{k}_{y} \\ 
0 & 0 & -\widehat{k}_{x}+i\widehat{k}_{y} & 0\ 
\end{array}%
\right) ,  \label{DH}
\end{equation}%
where $\mathrm{v}_{F}\approx c/300$ is the Fermi velocity ($c$ is the light
speed in vacuum). Note that $\hbar \widehat{\mathbf{k}}$\textbf{\ }is the
quasiparticle momentum operator and upper left (lower right) block of the
Hamiltonian (\ref{DH}) corresponds to $\mathbf{K}$ ($\mathbf{K}^{\prime }$)
point. In an armchair nanoribbon the wavefunction amplitude should vanish on
both sublattices at the extremes, $x=0$ and $x=W+a_{0}/2$, of the
nanoribbon. To satisfy this boundary condition one must admix valleys \cite%
{Brey}, and the confined wavefunctions have the form, 
\begin{equation}
\psi _{n,s,k_{y}}(\mathbf{r})=\frac{e^{ik_{y}y}}{2\sqrt{W+a_{0}/2}\sqrt{L_{y}%
}}\left( 
\begin{array}{c}
e^{-i\theta _{nk_{y}}}\,e^{ik_{n}x} \\ 
s\,e^{ik_{n}x} \\ 
e^{-i\theta _{nk_{y}}}\,e^{-ik_{n}x} \\ 
s\,e^{-ik_{n}x}%
\end{array}%
\right)  \label{free}
\end{equation}%
with energies 
\begin{equation}
\varepsilon _{n,s}(k_{y})=s\hbar \mathrm{v}_{F}\sqrt{k_{n}^{2}+k_{y}^{2}}
\label{energy}
\end{equation}%
{\ }for conduction ($s=1$) and valence ($s=-1$) bands. Here $\theta
_{nk_{y}}=\arctan k_{n}{/k_{y}}${. }Due to confinement in the $x$ direction
the allowed values of $k_{n}$ satisfy the quantization condition \cite{Brey} 
\begin{equation}
k_{n}=\frac{2\pi }{3a_{0}}+\frac{2\pi n}{2W+a_{0}}\,\,.  \label{kn}
\end{equation}%
For a width of the form $W\neq (3M+1)a_{0}$, nanoribbons have nondegenerate
states and are band insulators. The allowed values of $k_{n}$ are
independent of the momentum $k_{y}$.

\bigskip We will work in the second quantization formalism, expanding the
fermionic field operators on the basis of states given in (\ref{free}), that
is,%
\begin{equation}
\widehat{\Psi }(\mathbf{r})=\sum\limits_{n,s,k_{y}}\widehat{e}%
_{n,s,k_{y}}\psi _{n,s,k_{y}}(\mathbf{r}),  \label{exp}
\end{equation}%
where $\widehat{e}_{n,s,k_{y}}$ ($\widehat{e}_{n,s,k_{y}}^{\dagger }$) is
the annihilation (creation) operator for an electron. In (\ref{exp}) we have
omitted the real spin quantum number because of degeneracy. The total
Hamiltonian in the second quantization, reads: 
\begin{equation}
\hat{H}=\sum_{n,s,k_{y}}\mathcal{E}_{s,n,,k_{y}}\widehat{e}%
_{s,n,,k_{y}}^{\dag }\widehat{e}_{s,n,,k_{y}}+eE(t)\widehat{Y},  \label{H1}
\end{equation}%
where $e$ is the elementary charge and $\widehat{Y}$ is the second quantized
position operator along the $y$-direction. The latter can be expressed via
intraband ($\widehat{y}_{i}$) and interband ($\widehat{y}_{e}$) parts:%
\begin{equation*}
\widehat{Y}=\widehat{y}_{i}+\widehat{y}_{e}
\end{equation*}%
\begin{equation*}
\widehat{y}_{i}=i\sum\limits_{s,n,k_{y},k_{y}^{\prime }}\delta
_{k_{y}^{\prime }k_{y}}\partial _{k_{y}^{\prime }}\widehat{e}%
_{s,n,k_{y}}^{\dagger }\widehat{e}_{s,n,k_{y}^{\prime }}
\end{equation*}%
\begin{equation*}
\widehat{y}_{e}=\sum\limits_{n,k_{y}}\left( y_{\mathrm{tr}}\left(
n,k_{y}\right) \widehat{e}_{v,n,k_{y}}^{+}\widehat{e}_{c,n,k_{y}}+\mathrm{%
h.c.}\right) .
\end{equation*}%
Here%
\begin{equation}
y_{\mathrm{tr}}\left( n,k_{y}\right) =\langle -1,n,k_{y}|i\partial
_{k_{y}}|1,n,k_{y}\rangle =-\frac{1}{2}\frac{k_{n}}{k_{n}^{2}+k_{y}^{2}}.
\label{trdipm}
\end{equation}%
From the Heisenberg equation%
\begin{equation}
i\hbar \frac{\partial \widehat{e}_{\eta _{2},\mathbf{k}}^{\dag }\widehat{e}%
_{\eta _{1},\mathbf{k}}}{\partial t}=\left[ \widehat{e}_{\eta _{2},\mathbf{k}%
}^{\dag }\widehat{e}_{\eta _{1},\mathbf{k}},\widehat{H}\right] ,
\label{heis}
\end{equation}%
one can obtain the following evolution equations for the interband
polarization $\mathcal{P}_{n}(k_{y},t)=\langle \hat{e}_{1,n;k_{y}}^{+}\left(
t\right) \hat{e}_{-1,n;k_{y}}\left( t\right) \rangle $, and the distribution
functions for the conduction $\mathcal{N}_{c,n}\left( k_{y},t\right)
=\left\langle \hat{e}_{1,n;k_{y}}^{+}\left( t\right) \hat{e}%
_{1,n;k_{y}}^{+}\left( t\right) \right\rangle $ and valence $\mathcal{N}%
_{v,n}\left( k_{y},t\right) =\left\langle \hat{e}_{-1,n;k_{y}}^{+}\left(
t\right) \hat{e}_{-1,n;k_{y}}^{+}\left( t\right) \right\rangle $ bands%
\begin{equation*}
i\hbar \left[ \partial _{t}-eE_{y}\left( t\right) /\hbar \partial _{k_{y}}%
\right] \mathcal{P}_{n}(k_{y},t)+\left[ 2\hbar \mathrm{v}_{F}\sqrt{%
k_{n}^{2}+k_{y}^{2}}+i\hbar \Gamma _{n}\right] \mathcal{P}_{n}(k_{y},t)
\end{equation*}%
\begin{equation}
=-ey_{\mathrm{tr}}\left( n,k_{y}\right) E\left( t\right) \left(
N_{v,n}(k_{y},t)-N_{c,n}(k_{y},t)\right) ,  \label{pcv}
\end{equation}%
\begin{equation*}
i\hbar \left[ \partial _{t}-eE_{y}\left( t\right) /\hbar \partial _{k_{y}}%
\right] N_{c,n}(k_{y},t)+i\hbar \Gamma _{cn}\left(
N_{c,n}(k_{y},t)-N_{c,n}^{(0)}(k_{y})\right) 
\end{equation*}%
\begin{equation}
=ey_{\mathrm{tr}}\left( n,k_{y}\right) E\left( t\right) \mathcal{P}%
_{n}(k_{y},t)-\mathrm{c.c.},  \label{nc}
\end{equation}%
\begin{equation*}
i\hbar \left[ \partial _{t}-eE_{y}\left( t\right) /\hbar \partial _{k_{y}}%
\right] N_{v,n}(k_{y},t)+i\hbar \Gamma _{vn}\left(
N_{v,n}(k_{y},t)-N_{v,n}^{(0)}(k_{y})\right) 
\end{equation*}%
\begin{equation}
=-ey_{\mathrm{tr}}\left( n,k_{y}\right) E\left( t\right) \mathcal{P}%
_{n}(k_{y},t)-\mathrm{c.c.}.  \label{nv}
\end{equation}%
Where $\Gamma _{cn}$, $\Gamma _{vn}$, and $\Gamma _{n}$ are the
phenomenological relaxation rates which account for correlation terms
neglected in the free quasiparticle model. Here $N_{c,n}^{(0)}(k_{y})$ and $%
N_{v,n}^{(0)}(k_{y})$ are the equilibrium distribution functions to which
electrons and holes relax at rates $\Gamma _{cn}$ and $\Gamma _{vn}$,
respectively. For initial state, we assume Fermi-Dirac distribution: 
\begin{equation}
\mathcal{N}_{c,n}^{(0)}=\frac{1}{1+e^{\frac{\varepsilon
_{n,1}(k_{y})-\varepsilon _{F}}{T}}},\ \ \mathcal{N}_{v,n}^{(0)}=\mathcal{N}%
_{c,n}\left( -\varepsilon _{n,1}(k_{y})\right) ,\ \ \mathcal{P}%
_{n}(k_{y},t)=0.  \label{equ}
\end{equation}%
Here $\varepsilon _{F}$ is the Fermi energy and $T$ is the temperature. For
all calculations we assume the room temperature $T=0.025\,\mathrm{eV}$. The
dephasing rate $\Gamma _{n}$ in Eq. (\ref{pcv}) comprises all processes that
contribute to the decay of the interband polarization. In general $\Gamma
_{n}\geq \Gamma _{c,vn}$ and these rates vary with temperature and
quasiparticle density. Also, they have a quasiparticle momentum dependence,
which we presently ignore. In the extreme nonlinear response regime we will
assume that the main relaxation channel is the carrier--carrier collision on
the time scale of $50-100$ $\mathrm{fs}$ \cite{Binder,Malic}. Due to the
conservation of particles at the carrier--carrier collision: $\Gamma
_{n}=\left( \Gamma _{cn}+\Gamma _{vn}\right) /2$. Also taking into account
the electron-hole symmetry for the considered nanostructure, we assume $%
\Gamma _{cn}=\Gamma _{vn}$.

\section{Generation of harmonics}

We further examine the extreme nonlinear response of GNRs considering the
generation of harmonics at the multiphoton excitation. Nonlinear effects
take place when $eE_{0}y_{\mathrm{tr}}\left( n,k_{y}\right) $ becomes
comparable to or larger than photon energy $\hbar \omega $. Here we will
consider strong pump waves when $eE_{0}y_{\mathrm{tr}}\left( n,k_{y}\right)
>\hbar \omega $ for involved subbands. For the $10\,\mathrm{THz}$ photons
the nonlinear effects are essential already for a pump wave intensity $%
I_{0}=10^{7}\ \mathrm{W/cm}^{2}$. The pulse duration is taken to be $%
\mathcal{T}_{p}=40\mathcal{\pi }/\omega $. The integration of equations (\ref%
{pcv})-(\ref{nv}) is performed on a grid of $1000$ $k_{y}$-points
homogeneously distributed between the points $k_{\min }=-\alpha \omega /%
\mathrm{v}_{F}$ and $k_{\max }=\alpha \omega /\mathrm{v}_{F}$, where $\alpha 
$ depends on the intensity of the pump wave. Then we take into account $5$
subbands in our calculation. The time integration is performed with the
standard fourth-order Runge-Kutta algorithm.

The optical excitation via coherent radiation pulse creates electron-hole
pairs which result in the macroscopic current, providing two sources%
\begin{equation}
j_{y}\left( t\right) =j_{y\mathrm{e}}\left( t\right) +j_{y\mathrm{a}}\left(
t\right)  \label{cur}
\end{equation}%
for the generation of harmonics radiation. The first term in Eq. (\ref{cur}%
), which can be written by means of polarization,%
\begin{equation}
j_{y\mathrm{e}}\left( t\right) =\ \frac{ig_{s}e}{\hbar W}\left\langle \left[ 
\widehat{y}_{e},\widehat{H}_{0}\right] \right\rangle =-\frac{2e}{\hbar W}%
\sum_{n,k_{y}}\left( \frac{2\mathcal{E}_{c,n,k_{y}}}{i}y_{\mathrm{tr}}\left(
n,k_{y}\right) \mathcal{P}_{n}^{\ast }(k_{y},t)+\mathrm{c.c.}\right) ,
\label{inter}
\end{equation}%
is the interband current, and the second term, which is defined via
distribution functions,%
\begin{equation}
j_{y\mathrm{a}}\left( t\right) =\frac{ig_{s}e}{\hbar W}\left\langle \left[ 
\widehat{y}_{i},\widehat{H}_{0}\right] \right\rangle =-\frac{2e}{\hbar W}%
\left[ \sum\limits_{n,k_{y}}\partial _{k_{y}}\mathcal{E}%
_{c,n,k_{y}}N_{c,n}(k_{y},t)+\partial _{k_{y}}\mathcal{E}%
_{v,n,k_{y}}N_{v,n}(k_{y},t)\right]  \label{intra}
\end{equation}%
is the intraband current. Here $g_{s}=2$\ is the spin degeneracy factor. The
HHG spectrum is obtained from the Fourier transform $E^{(g)}(\omega )$ of
the function $E^{(g)}\left( t\right) =4\pi j_{y}\left( t\right) /c$, which
is the generated electrical field for 2D patterned graphene nanostructure.

As was mentioned in the previous section, the energy spectrum of
quasiparticles strongly depends on the width of GNRs. For a width of the
form $W=(3M+1)a_{0}$, nanoribbons are metallic, otherwise GNRs are band
insulators. Thus we have made calculation for both cases and the typical HHG
spectra are shown in Fig. 1 where we plot the HHG yield via logarithm of the
radiation intensity $c\left\vert E^{(g)}\left( \omega \right) \right\vert
^{2}/4\pi \ $for the GNRs at the various widths $W=a_{0}N$. In Figs. 1-4 for
the relaxation rates we assume $\hbar \Gamma _{n}=\hbar \Gamma _{c,vn}=0.05\,%
\mathrm{eV}$. From Fig. 1(a) we see considerable enhancement of the HHG
yield up to the middle of the spectra for metallic GNR. For metallic case as
in the graphene we have gapless spectrum which causes effective creation of
electron-hole pairs, electron-hole acceleration, and recollision with
emission of harmonics. There is no sharp cutoff of harmonics. In contrast to
metallic one for band insulator ($N=17$) we have plateau in the HHG spectrum
with the sharp cutoff. For large widths $N=40$ and $41$ the difference
between the both cases is minimal since energy gap becomes smaller than the
Fermi energy.

\begin{figure}[tbp]
\includegraphics[width=.9\textwidth]{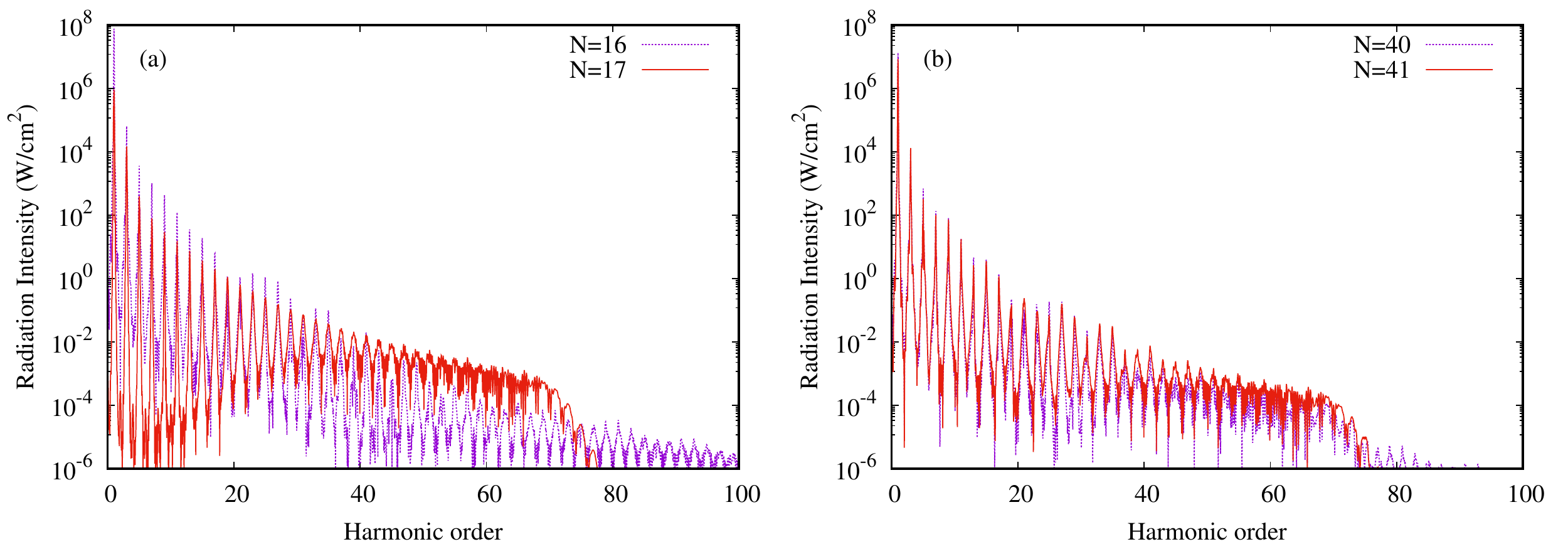}
\caption{The HHG yield via logarithm of the radiation intensity $c\left\vert
E^{(g)}\left( \protect\omega \right) \right\vert ^{2}/4\protect\pi \ $for
the GNRs with low Fermi energy $\protect\varepsilon _{F}=0.1\,\mathrm{eV}$
at the various widths $W=a_{0}N$, which define the single particle spectrum.
(a) for $N=16$ (metallic) and $N=17$ (insulator) and (b) for larger widths: $%
N=40$ (metallic) and $N=41$ (insulator). We assume $\hbar \protect\omega %
=0.041\,\mathrm{eV}$ ($\protect\omega /(2\protect\pi )=10\ \mathrm{THz}$).
The pump wave intensity is taken to be $I_{0}=5.0\times 10^{8}\ \mathrm{W/cm}%
^{2}$.}
\label{s2}
\end{figure}

\begin{figure}[tbp]
\includegraphics[width=.9\textwidth]{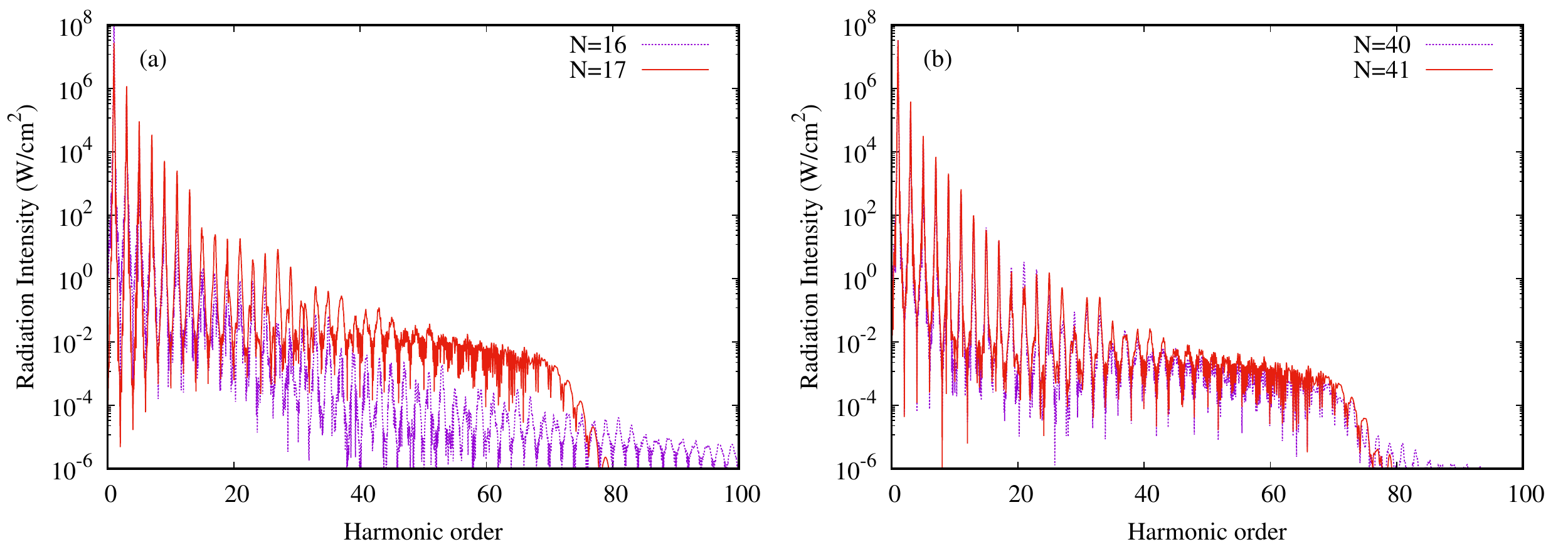}
\caption{The same as for Fig. 1 but for large Fermi energy $\protect%
\varepsilon _{F}=0.4\,\mathrm{eV}$.}
\end{figure}

In Fig. 2 we plot HHG spectra for relatively large Fermi energy $\varepsilon
_{F}=0.4\,\mathrm{eV}$\textrm{.} In this case the situation is opposite.
From Fig. 2(a) we see considerable enhancement of the HHG yield for the
entire spectra for nonmetallic GNR. This is connected with the Pauli
blocking. Thus, for large $\varepsilon _{F}>>\hbar \omega $ Pauli blocking
reduces the probability of creation of electron-hole pairs in the case of
gapless quasienergy spectrum. As in the case of Fig. 1(b) for relatively
large widths the difference between both cases is minimal (see Fig. 2(b)).

We have also investigated the intensities of 3rd, 5th, 7th and 9th harmonics
versus GNR width for Fermi energies $\varepsilon _{F}=0.1\,\mathrm{eV}$ and $%
\varepsilon _{F}=0.4\,\mathrm{eV}$. The latter is plotted in Fig. 3(a) and
3(b). For visual convenience we have rescaled the intensities. As is seen
from Fig. 3(a), the intensities for the small Fermi energy are maximal in
metallic GNRs ($N=13,16,19...$) up to $N=28$ harmonics. From Fig 3(b), we
see that for the large Fermi energies overall the intensities are maximal in
nonmetallic GNRs.

\begin{figure}[tbp]
\includegraphics[width=.9\textwidth]{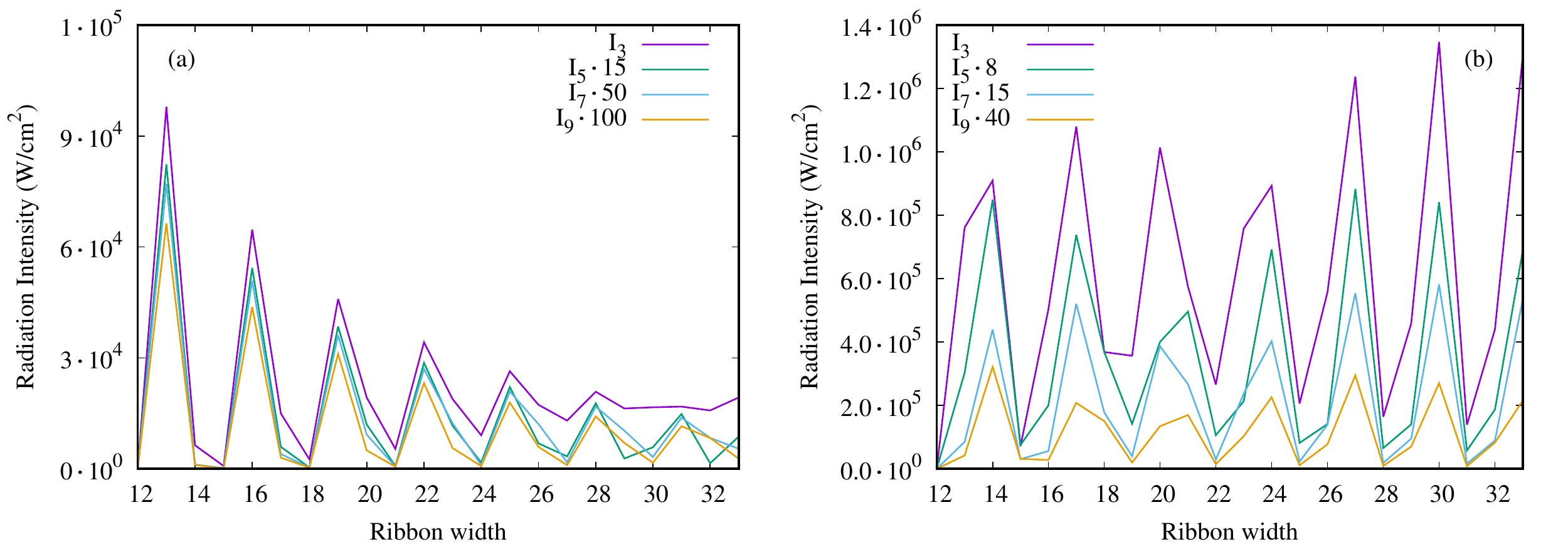}
\caption{The intensities of 3rd, 5th, 7th and 9th harmonics versus
nanoribbon width in units of $a_{0}$ ($N=W/a_{0}$). We assume $\protect%
\omega /(2\protect\pi )=10\ \mathrm{THz}$ and $I_{0}=5.0\times 10^{8}\ 
\mathrm{W/cm}^{2}$: (a) $\protect\varepsilon _{F}=0.1\,\mathrm{eV}$ and (b) $%
\protect\varepsilon _{F}=0.4\,\mathrm{eV}$.}
\end{figure}

As we see from Figs. 1 and 2, the high-order harmonics up to the 80th orders
are appeared. Note that only odd harmonics are generated, reflecting the
inversion symmetry preserved in the GNRs. One of the main questions at HHG
is the cutoff harmonic dependence on the intensity of the pump wave. In Fig.
4, we plot the HHG yield for the GNR of width $W=17a_{0}$ for the various
pump wave intensities at fixed frequency. As is seen, the cutoff harmonic is
proportional to $I^{1/2}$. 
\begin{figure}[tbp]
\includegraphics[width=.45\textwidth]{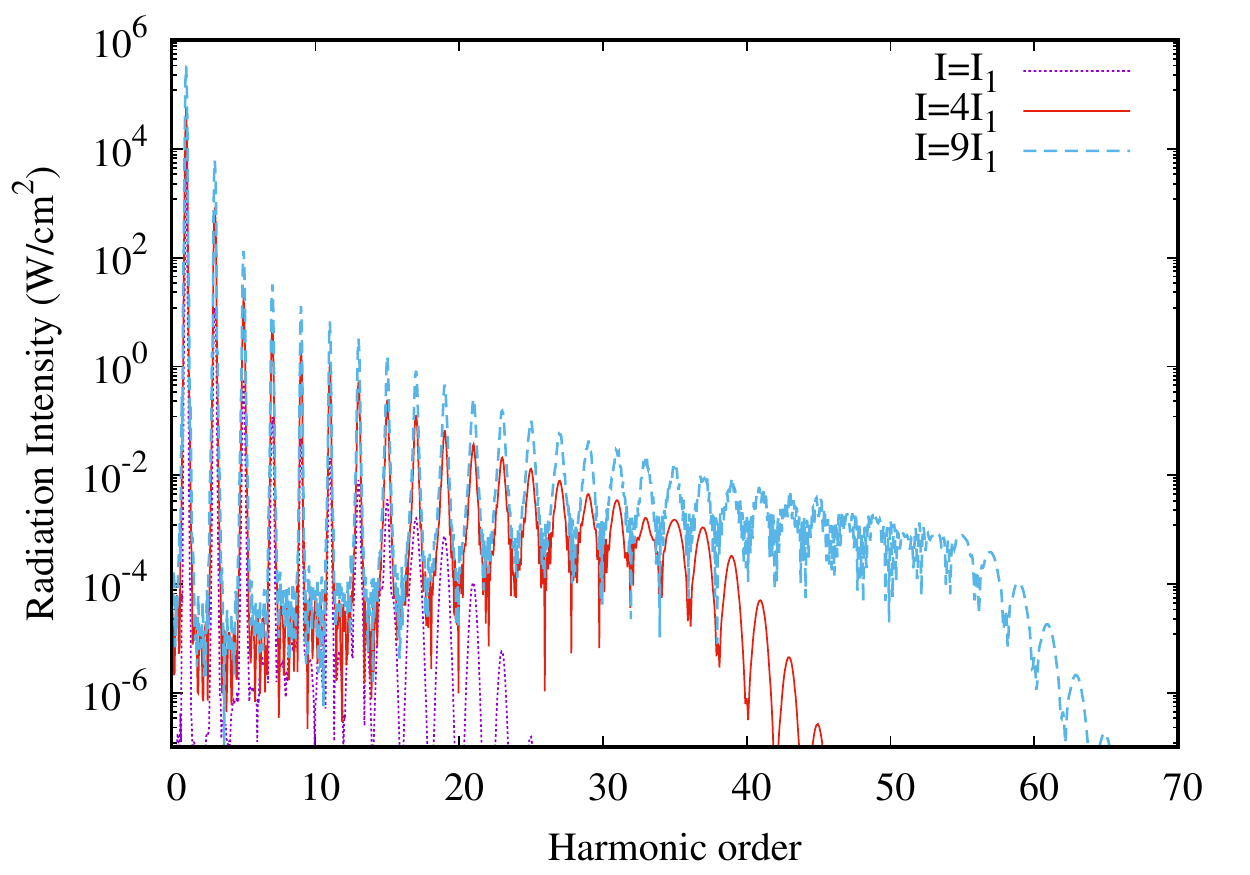}
\caption{The HHG yield for the GNR of width $W=17a_{0}$ for various pump
wave intensities at fixed frequency. We assume $\hbar \protect\omega =0.041\,%
\mathrm{eV}$ ($\protect\omega /(2\protect\pi )=10\ \mathrm{THz}$), $%
I_{1}=3.5\times 10^{7}\ \mathrm{W/cm}^{2}$ and Fermi energy $\protect%
\varepsilon _{F}=0.1\,\mathrm{eV}$\textrm{.}}
\end{figure}
In order to see the physical origin behind this dependence we will examine
Eq. (\ref{pcv}). Formally the solution of the latter can be written as 
\begin{equation*}
\mathcal{P}_{n}(k_{y},t)=\frac{ie}{\hbar }\int_{0}^{t}dt^{\prime }y_{\mathrm{%
tr}}\left( n,\widetilde{k}\left( t,t^{\prime }\right) \right) E\left(
t^{\prime }\right) \left( N_{v,n}(\widetilde{k}\left( t,t^{\prime }\right)
,t^{\prime })-N_{c,n}(\widetilde{k}\left( t,t^{\prime }\right) ,t^{\prime
})\right)
\end{equation*}%
\begin{equation}
\times \exp \left( 2i\mathrm{v}_{F}\int_{t^{\prime }}^{t}\sqrt{k_{n}^{2}+%
\widetilde{k}\left( t,t_{1}\right) }dt_{1}\right) \exp \left( -\Gamma
_{n}\left( t-t^{\prime }\right) \right)  \label{sol}
\end{equation}%
where 
\begin{equation*}
\widetilde{k}\left( t,t^{\prime }\right) =k_{y}+\frac{e}{\hbar }%
\int_{t^{\prime }}^{t}E\left( \tau \right) d\tau
\end{equation*}%
is the classical momentum change in the wave field. The time dependence of $%
\mathcal{P}_{n}(k_{y},t)$ is mainly determined by the exponential factor
with the electron-hole energy in the field $\ \varepsilon _{eh}\left(
n,t,t^{\prime }\right) =2\mathrm{v}_{F}\hbar \sqrt{k_{n}^{2}+\widetilde{k}%
\left( t,t^{\prime }\right) }$. The cutoff frequency is determined by
electron-hole pairs recolliding with the highest energy $\hbar \omega
_{c}\sim \varepsilon _{eh}\left( n,t,t^{\prime }\right) _{\max }$. For the
strong wave fields $\widetilde{k}_{\max }=2eE_{0}/\hbar \omega >>k_{n}$ this
yields to the linear dependence of the cutoff frequency on the pump
radiation field: $\hbar \omega _{c}\simeq 4\mathrm{v}_{F}eE_{0}/\omega $.
This cutoff frequency is close to numerical values determined from Fig. 4.

\begin{figure}[tbp]
\includegraphics[width=.9\textwidth]{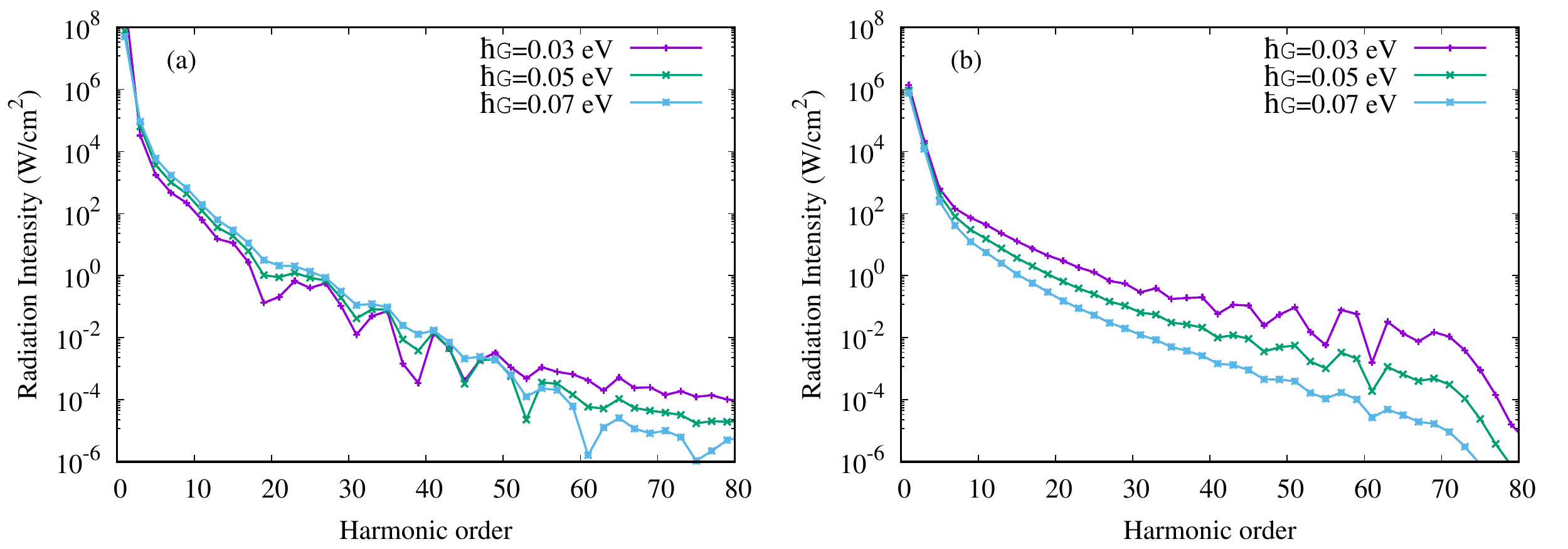}
\caption{The HHG yield for the GNR for various relaxation rates. We assume $%
\protect\omega /(2\protect\pi )=10\ \mathrm{THz}$, $I_{0}=5.0\times 10^{8}\ 
\mathrm{W/cm}^{2}$, and $\protect\varepsilon _{F}=0.1\,\mathrm{eV}$: (a) $%
N=W/a_{0}=16$ and (b) $N=17$.}
\end{figure}

We have also investigated the HHG yield at various relaxation rates ($\Gamma
\equiv \Gamma _{n}$) for metallic ($N=16$) and for band insulator ($N=17$)
GNRs. The latter is plotted in Fig. 5. For visual convenience in the
logarithmic scale we have plotted the envelope of the intensities on $2s+1$
harmonics . In Fig. 5 we see the considerable difference of the HHG yield
depending on the quasiparticle spectrum of GNR. Fig. 5(a) shows the
robustness of HHG in metallic GNR against relaxation processes in contrast
to a band insulator case demonstrated in Fig. 5(b), where harmonics are
suppressed at high relaxation rates.

\section{Conclusion}

We have presented the microscopic theory of nonlinear interaction of the
GNRs with a strong coherent radiation field. For the extreme nonlinear
optical response, we have used a free quasiparticle model and obtained a
closed set of differential equations for the single-particle density matrix
with the phenomenological relaxation terms. These equations have been solved
numerically. We have considered multiphoton excitation of GNRs towards the
high-order harmonics generation. It has been shown that the width size and
Fermi energy level of the GNR in the nonlinear optical response are quite
considerable. For the low Fermi energies $\varepsilon _{F}\sim \hbar \omega $
we obtained a considerable enhancement of the HHG yield up to the middle of
the spectra in metallic GNR compared with the nonmetallic ones. For the
large Fermi energies $\varepsilon _{F}>>\hbar \omega $, the nonmetallic GNRs
are more effective for HHG. We extracted the linear upon the pump field
amplitude dependence on the HHG cutoff frequency. Obtained results show that
GNRs can serve as an effective medium for the high-order harmonic generation
with radiation fields of moderate intensities due to the confinement of
quasiparticles in GNRs.

\begin{acknowledgments}
This work was supported by the RA MES State Committee of Science and
Belarusian Republican Foundation for Fundamental Research (RB) in the frames
of the joint research project SCS 18BL-020.
\end{acknowledgments}

\end{document}